\documentclass[iop]{emulateapj}
\usepackage{epstopdf}
\usepackage{color}
\shorttitle{Constraints on Lorentz Invariance Violation with gamma-ray bursts via a Markov Chain Monte Carlo approach}
\shortauthors{Pan et al.}

\newcommand{\beq}{\begin{equation}}
\newcommand{\eeq}{\end{equation}}

\def\ba{\begin{eqnarray}}
\def\ea{\end{eqnarray}}

\begin{document}
\title{Constraints on Lorentz Invariance Violation with gamma-ray bursts via a Markov Chain Monte Carlo approach}
\author{Yu Pan\altaffilmark{1,2}, Yungui Gong\altaffilmark{3}, Shuo Cao\altaffilmark{2}, He Gao\altaffilmark{4}and Zong-Hong Zhu\altaffilmark{2*}
}

\altaffiltext{1}{College of Science, Chongqing
University of Posts and Telecommunications, Chongqing 400065, China }
\altaffiltext{2}{Department of Astronomy, Beijing Normal University, Beijing 100875,
China; \emph{zhuzh@bnu.edu.cn}}
\altaffiltext{3}{MOE Key Laboratory of Fundamental Quantities Measurement, School of Physics, Huazhong University of Science and Technology, Hubei 430074, China}
\altaffiltext{4}{Department of Astronomy and Astrophysics, Pennsylvania State University, 525 Davey
 Laboratory, University Park, PA 16802}

\begin{abstract}

In quantum theory of gravity, we expect the Lorentz Invariance
Violation (LIV) and the modification of the dispersion relation
between energy and momentum for photons. The effect of the
energy-dependent velocity due to the modified dispersion relation
for photons was studied in the standard cosmological context by
using a sample of Gamma Ray Bursts (GRBs). In this paper we mainly
discuss the possible LIV effect by using different cosmological
models for the accelerating universe. Due to the degeneracies among
model parameters, the GRBs' time delay data are combined with the
cosmic microwave background data from the
Planck first year release, the baryon acoustic oscillation
data at six different redshifts, as well as Union2 type Ia
supernovae data, to constrain both the model parameters and the LIV
effect. We find no evidence of LIV.

\end{abstract}

\section{Introduction}

Above the Planck energy scale $E_{\rm Pl}$, we expect a quantum theory of gravity in place of
Einstein's general relativity.
The quantization of space-time will lead to the modification of the dispersion relation between energy and momentum
of a particle with mass $m$ and the breakdown of Lorentz invariance.
In some quantum gravity models, the consequence of the Lorentz Invariance Violation (LIV)
is the energy dependence of the speed of light in vacuum \citep{Amelino-Camelia98,Amelino-Camelia01}.
In these scenarios, the energy dependent velocity of light is $v=c(1-\xi E/E_{QG})$ with effective quantum gravity energy scale $E_{QG}$,
the speed of high energy photons is slower, and low energy photons will reach us earlier. Therefore,
the measurement of the light speed in vacuum can be used to study the LIV effect.

In the past years, both astrophysics
and particle physics' experiments have been
proposed to test the LIV and quantum gravity effects
\citep{Mattingly05,Sarkar02,Amelino-Camelia13}. Because photons
with different energy will reach us at different time, and the
energy of Gamma-Ray Bursts (GRBs) which are the most luminous
explosions in the universe with short duration ranges from KeV to
GeV, GRBs with intense bursts of $\gamma$-ray photons from
cosmological distances are proposed to
 measure the difference of the arrival time of photons with different energy \citep{Amelino-Camelia98,Amelino-Camelia02,Kowalski-Glikman02}.
GRBs were also used to constrain LIV theories \citep{Ellis03, Ellis06,Rodriguez06,Jacob07}.
\citet{Biller99} used the data for
a TeV $\gamma$-ray flare coming from the active galaxy Markarian 421 to
set a bound on the energy scale of quantum gravity as $E_{QG}>6\times 10^{16}$ GeV.
\citet{abdo09a} used GRB 090510 to study the LIV effect and get $E_{QG}>1.2 E_{\rm Pl}$,
the energy scale of quantum gravity was found to be $E_{QG}>1.3\times 10^{18}$ GeV by using GRB 080916C.
The LIV effects due to extra dimensions can also be constrained by GRBs \citep{Baukh07}.
Experiments with cold atoms were also proposed to constrain the modified dispersion
relation due to quantum gravity \citep{Amelino-Camelia09}.

In order to measure the LIV, statistical and possible systematic uncertainties must be minimized.
For this purpose, \citet{Ellis03,Ellis06} developed a method to analyze samples of GRBs with different
redshifts and energy bands. The technique has the advantage of extracting time-dependent features from the signals of many GRBs.
In the analysis \citep{Ellis03}, they used both the BATSE and OSSE data from the Compton Gamma Ray Observatory \footnote{ftp://legacy.gsfc.nasa.gov/compton/data/batse/ascii/data/64ms/}.
By adding larger samples of GRBs with known redshifts from HETE \footnote{http://space.mit.edu/HETE/Bursts/} and
SWIFT \footnote{http://swift.gsfc.nasa.gov/docs/swift/archive/},
the observed time delay was formulated in terms of linear regression where the intercept denotes intrinsic time delay and
the linear term denotes LIV effect \citep{Ellis08,Jacob08}. By using the concordance $\Lambda$CDM model with
$\Omega_{\Lambda}=0.7$, they found no strong evidence of
LIV, and they obtained $E_{QG} \geq 1.4 \times 10^{16}$ GeV \citep{Ellis06}.

Since the distances from the sources to the observer depend on
cosmological models, the conclusions on the LIV and quantum gravity
effect may be affected by the cosmological model used in the
analysis. However, only the concordance $\Lambda$CDM scenario
($\Omega_{\Lambda}=0.7$, $\Omega_m=0.3$) was studied
\citep{Ellis03}. Therefore, it is necessary to consider other
cosmological models. Recently \citet{Biesiada09} studied the LIV by
using the quintessence and Chaplygin Gas model and they found a weak
evidence of LIV. In their analysis, the degeneracies among model
parameters were neglected, as they were taken as fixed values. In
this paper we mainly discuss the possible LIV by using different
cosmological models which explain the present cosmic acceleration.
Compared with the previous work, in addition to the relevant
parameters quantifying LIV, the cosmological parameters are also
treated as free parameters. We adopt the Markov Chain Monte
Carlo (MCMC) technique to constrain LIV and model parameters with
the observational data. In order to derive tighter constraints on the model parameters,
we combine the time delay data from GRBs with the cosmic microwave background (CMB) data from the
Planck first year release \citep{Planck1,Planck2}, the baryon
acoustic oscillation (BAO) data
\citep{Blake11b,Beutler11,Gong13,Percival10}, and the 557 Union2
SNeIa data \citep{Amanullah10}. This paper is organized as
follows: we introduce the LIV in section II. The GRBs' time
delay and other observational data are discussed in section III. The
constraint results on LIV and cosmological parameters with different
cosmological models are presented in section IV, and finally, the
conclusions are drawn in section V.


\section{The Lorentz Invariance Violation } \label{sec:Violation}

The modified dispersion relation due to quantum gravity models can
be parameterized as follows \citep{Ellis03,Biesiada09}
\begin{equation}
E^2-p^2c^2=\epsilon{E^2}\left(\frac{E}{\xi_{n}E_{QG}}\right)^n,
\end{equation}
where $\xi_n$ is a dimensionless parameter,
$\xi_1=1$, $\xi_2=10^{-7}$ \citep{Jacob07}, $\epsilon=+1$,
and the effective energy scale $E_{QG}$ of quantum gravity
is expected to be near the Planck scale.
Because the LIV effects are small, \citep{Ellis08,Biesiada09},
we consider the $n=1$ term only in this study.

Since the arrival time for photons with
energy $E$ is equal to \citep{Jacob07,Jacob08,Biesiada07,Biesiada09}
 \begin{equation}
t_{LIV} = \int_{0}^{z} \left[ 1 + \frac{E}{E_{QG}} (1+z')
\right] \frac{dz'}{H(z')},
\label{flight time}
\end{equation}
so the the arrival times of two photons with different energy will be different,
and the difference in time for the energy difference $\Delta E$ is
\begin{equation}
\Delta t_{LIV} = \frac{\Delta{E}}{H_0 E_{QG}}
  \int_0^z  \frac{(1+z')dz'}{h(z')},
\label{time delay}
\end{equation}
where the dimensionless Hubble parameter $h(z) = H(z)/H_0$,
$H_0$ is the Hubble constant and $H(z)$ is the Hubble parameter at redshift $z$.
In order to account for the unknown intrinsic time lags,
we fit the measured time-lags by including a parameter $b$ specified
in the rest frame of the source. The arrival time delays are fitted by the formula
$\Delta t_{obs} = \Delta t_{LIV} + b(1+z)$ \citep{Ellis06}.
Therefore, the simple linear fitting function is
\begin{equation}
\frac{\Delta t_{obs}}{1+z} = a_{LIV}K + b,
\end{equation}
where the intercept represents the intrinsic time lags, the LIV effects are encoded in the slope
$a_{LIV}= \Delta E/(H_0E_{QG})$ which is related to the scale of Lorentz violation \citep{Jacob08, Biesiada07} and
$K=(1+z)^{-1} \int^z_0 dz' (1+z')/h(z')$ is related to the measurements of cosmic distances. The model dependence
is through the function $K(z)$ which will be calculated for three popular cosmological models.

\section{The Observational data }

In this paper we use the BATSE data with 9 light curves
whose time resolution is 64 ms and redshifts span from from $z=0.835$ to $z=3.9$,
the HETE data with 15 light curves whose time resolution is 164 ms
and redshifts span from $z=0.168$ to $z=3.372$,
and the SWIFT data with 11 light curves whose time resolution is 64 ms
and redshifts span from $z=0.258$ to $z=6.29$, the data was shown in Table 1 of \citet{Ellis06}.
So we use the data of time lags between different energy channels measured from the light curves
of 35 GRBs with redshifts from $z=0.168$ to $z=6.29$ \citep{Ellis06}.
The spectral time lags are obtained from the light curves
in the 115-320 keV energy band with respect to those in the lowest 25-55 keV energy band.
To test the LIV effect,
we apply the $\chi^2$ statistics, here $\chi^2$ is
\begin{equation}
\chi^2_{GRBs}=\sum_{i=1}^{N_{GRBs}}\left[\frac{\Delta{t_i}-\Delta {t_{obs}}}{\sigma_{i}}\right]^2,
\end{equation}
where $\Delta t_i$ and $\Delta {t_{obs}}$ respectively
denote the theoretical and observational values of time delays of
GRBs, and $\sigma_i$ is the observational uncertainty.

In fitting the data, in addition to the slope $a_{LIV}$ and the intercept $b$, we also need to
fit the model parameters. In previous studies, the model parameters are fixed and
the degeneracies among parameters are neglected. To account for the degeneracies among these parameters,
it would be better to treat the model parameters as free parameters and estimate their nominal values
from the observational data.
With this aim, we combine the CMB data from the Planck first year release, the 557 Union2 SNeIa data,
and the BAO data with the data of spectral time lags from GRBs to constrain the model parameters.
Since we have at least four parameters to fit, we
take the MCMC \citep{Lewis02} technique to constrain the model parameters.

For the 557 Union2 SNeIa data, the distance modulus $\mu(z)$ is measured at different redshifts, and
the theoretical value of the distance modulus is
\begin{equation}
\mu=5\log\frac{d_L}{Mpc}+25=5\log_{10}H_0d_L-\mu_0,
\end{equation}
where $\mu_0=5\log_{10}[H_0/(100\ {\rm km/s/Mpc})]-42.38$, and the luminosity distance is
$d_L=(1+z)H_0^{-1}\int_0^z dz'/E(z')$. To fit the 557 Union2 SNeIa data, we calculate
\begin{equation}
\chi^2_{SNe}=\sum_{i=1}^{N}\frac{[\mu(z_i)-\mu_{obs}(z_i)]^2}{\sigma_{\mu{i}}^2},
\end{equation}
where $\mu(z_i)$ and $\mu_{obs}(z_i)$ are the theoretical and observed distance modulus for the SNeIa at redshift
$z_i$. The $\sigma_{\mu{i}}$ is observational error.
For the nuisance parameter $H_0$, we marginalize it with a flat prior.

The BAO $A$ data consist of the measurements of $A$ at three redshifts $z=0.44$, $z=0.6$ and $z=0.73$ from WiggleZ dark energy survey \citep{Blake11b,Gong13}, and the parameter $A$ is defined as
\begin{equation}
A=\sqrt{\Omega_{m}}\frac{H_0 D_V(z)}{z},
\end{equation}
where the effective distance is given by \citep{Eisenstein05}
\begin{equation}
\label{dvdef}
D_V(z)=\left[\frac{d_L^2(z)}{(1+z)^2}\frac{z}{E(z)}\right]^{1/3}.
\end{equation}
The $\chi^2$ value for this data set is
\begin{eqnarray}
\chi^2_{\mathrm{BAO1}}=\Delta
\bf{A}_{\mathrm{BAO}}^\mathrm{T}{\bf
C_{\mathrm{BAO}}}^{-1}\Delta\bf{A}_{\mathrm{BAO}},
\end{eqnarray}
where ${\bf{C_{\mathrm{BAO}}}^{-1}}$ is the corresponding inverse covariance matrix (See Table~\ref{tabwigcov}).

\begin{table}
\caption{The corresponding inverse covariance matrix of $A$. \label{tabwigcov}}
\begin{center}
\begin{tabular}{cccc}
\hline
Redshift slice & $0.2<z<0.6$ & $0.4<z<0.8$ & $0.6<z<1.0$ \\
\hline
$0.2<z<0.6$ & $1040.3$ & $-807.5$ & $336.8$ \\
$0.4<z<0.8$ & & $3720.3$ & $-1551.9$ \\
$0.6<z<1.0$ & & & $2914.9$ \\
\hline
\end{tabular}
\end{center}
\end{table}

In addition to the above $A$ parameter, we also consider the BAO distance
ratio $d_z$ at $z=0.2$ and $z=0.35$ from SDSS data release 7 (DR7) galaxy sample \citep{Percival10}
and $d_{0.106}=0.336\pm{0.015}$ from the 6dFGS measurements \citep{Beutler11}.
The BAO distance ratio is
\begin{equation}
\label{dz} d_{z}= \frac{r_{s}(z_{d})}{D_{V}(z)},
\end{equation}
where the comoving sound horizon is
\begin{equation}
\label{rshordef} r_s(z)=\int_z^\infty \frac{c_s(z)dz}{E(z)},
\end{equation}
the sound speed $c_s(z)=1/\sqrt{3[1+\bar{R_b}/(1+z)}]$, and $\bar{R_b}=3\Omega_b h^2/(4\times2.469\times10^{-5})$.
As usual, we fit the drag redshift $z_d$ as follows \citep{Eisenstein98}
\begin{eqnarray}
\label{zdfiteq}
z_d&=&\frac{1291(\Omega_m h^2)^{0.251}}{1+0.659(\Omega_m h^2)^{0.828}}[1+b_1(\Omega_b
h^2)^{b_2}], \nonumber\\
b_1&=&0.313(\Omega_m h^2)^{-0.419}[1+0.607(\Omega_m
h^2)^{0.674}],\nonumber\\
b_2&=&0.238(\Omega_m h^2)^{0.223}.
\end{eqnarray}

The $\chi^2$ value of the BAO distance ratio is \citep{Percival10}
\begin{eqnarray}
\chi^2_{\mathrm{BAO2}}=\Delta
\bf{P}_{\mathrm{BAO}}^\mathrm{T}{\bf
C_{\mathrm{BAO}}}^{-1}\Delta\bf{P}_{\mathrm{BAO}},
\end{eqnarray}
where $\bf{\Delta{P}_{BAO}}=\bf{P}_{th}-\bf{{P}_{obs}}$, $\bf{{P}_{obs}}$ is the observed distance ratio,
and ${\bf{C_{\mathrm{BAO}}}}$ is the covariance matrix for the distance ratio.
To use the BAO data, we minimize
\begin{equation}
\chi^2_{\mathrm{BAO}}=\chi^2_{\mathrm{BAO1}}+\chi^2_{\mathrm{BAO2}}+\frac{(d_{0.106}-0.336)^2}{0.015^2}.
\end{equation}

For the CMB measurement, we use the derived parameters including the acoustic scale $l_a$,
the shift parameter $R$, and $\Omega_bh^2$ \citep{Planck1,Planck2}
from the Planck temperature measurements combined with lensing,
as well as WMAP polarization data at low multipoles with $l \leq 23$.
The acoustic scale is
\begin{equation}
l_a=\pi\frac{\Omega_\mathrm{k}^{-1/2}H_0^{-1}{\rm sinn}[\Omega_\mathrm{k}^{1/2}\int_0^{z_{\ast}} dz/E(z)]}{r_s(z_{\ast})},
\end{equation}
where $r_s(z_{\ast})
={H_0}^{-1}\int_{z_{\ast}}^{\infty}c_s(z)/E(z)dz$ is the comoving sound horizon at the recombination.
The shift parameter is
\begin{equation}
R(z_\ast)=\frac{\sqrt{\Omega_{m}}}{\sqrt{|\Omega_{k}|}}{\rm
sinn}\left(\sqrt{|\Omega_{k}|}\int_0^{z_\ast}\frac{dz}{E(z)}\right).
\end{equation}
The recombination redshift $z_\ast$ is fitted by \citep{Hu96}
\begin{eqnarray}
\label{zasteq} z_\ast&=&1048[1+0.00124(\Omega_b
h^2)^{-0.738}][1+g_1(\Omega_m h^2)^{g_2}], \nonumber\\
g_1&=&\frac{0.0783(\Omega_b h^2)^{-0.238}}{1+39.5(\Omega_b
h^2)^{0.763}},\nonumber\\
g_2&=&\frac{0.560}{1+21.1(\Omega_b h^2)^{1.81}}.
\end{eqnarray}

The $\chi^2$ value for the CMB data is
\begin{equation}
\chi^2_{\mathrm{CMB}}=\Delta
\textbf{P}_{\mathrm{CMB}}^\mathrm{T}{\bf
C_{\mathrm{CMB}}}^{-1}\Delta\textbf{P}_{\mathrm{CMB}},
\end{equation}
where $\Delta\bf{P_{\mathrm{CMB}}} =
\bf{P_{\mathrm{th}}}-\bf{{P}_{\mathrm{obs}}}$, $\bf{{P}_{\mathrm{obs}}}$ and $\bf{{P}_{\mathrm{th}}}$
are the observed and the theoretical values of the derived parameters, respectively, and
${\bf C_{\mathrm{CMB}}}$  is the  covariance matrix  for the derived CMB data \citep{Planck2}.

\begin{table}
\begin{center}
\caption{The $1\sigma$ and $2\sigma$ constraints on $a$ and $b$ for different cosmological model.}
\begin{tabular}{|c|c|} \hline
Cosmological model&Regression coefficient $a$  \\
                  &  and intercept $b$ \\
\hline
$\Lambda$CDM & $a=-0.017_{-0.0718,-0.1415}^{+0.0717,+0.1416}$ \\
             & $b=-0.00013_{-0.0155,-0.0305}^{+0.0154,+0.0308}$ \\
\hline
$w$CDM & $a=-0.0168_{-0.0702,-0.1392}^{+0.0711,+0.1397}$  \\
       & $b=-0.00015_{-0.0154,-0.0304}^{+0.0153,+0.0303}$ \\
\hline
CPL & $a=-0.0183_{-0.0711,-0.1401}^{+0.0712,+0.14}$ \\
      & $b=0.00018_{-0.0155,-0.0301}^{+0.0159,+0.0306}$ \\
\hline
\end{tabular}\label{all}
\end{center}
\end{table}

\section{Constraints on the LIV parameters and analysis} \label{sec:model}

To explain the cosmic acceleration,
an exotic energy component called dark energy with
negative pressure was proposed \citep{Riess98,Perlmutter99,Astier06,Hicken09,Amanullah10,Spergel03,Spergel07,Komatsu09,Komatsu11,Tegmark04,Eisenstein05,Cao11,Cao12a,Cao12b,Gao14,Gong12,Gong13}.
The most simple candidate for dark energy is the vacuum energy known as the cosmological constant $\Lambda$.
Because of the huge difference between predicted and measured values
of the vacuum energy, many other dynamical dark energy models have also been considered, including quintessence \citep{Ratra88, Caldwell98}, phantom \citep{Caldwell02,Caldwell03}, k-essence \citep{Armendariz-Picon01,Chiba02}, as well as quintom models \citep{Feng05,Feng06,Guo05}.
In this paper, we consider three different dark energy models: the $\Lambda$CDM model, the dark energy model with constant
equation of state parameter $w$, and the Chevallier-Polarski-Linder (CPL) model \citep{Chevallier01,Linder03}.

\begin{figure}
\begin{center}
\includegraphics[width=0.9\hsize]{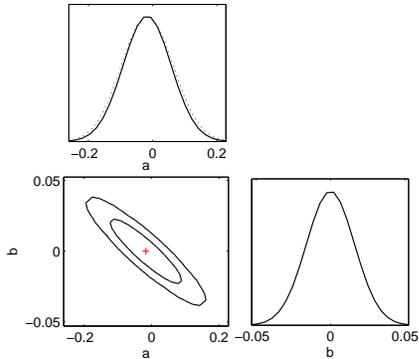}
\end{center}
\caption{ The marginalized 1$\sigma$ and 2$\sigma$ contours and the distribution of the LIV parameters for the $\Lambda$CDM model. The red cross represents the best-fit point.
\label{L1}}
\end{figure}

For the $\Lambda$CDM model, the equation of state parameter $w=p/\rho=-1$ and the Friedmann equation is
\begin{equation}
h^2(z)=\Omega_m(1+z)^3 + \Omega_\Lambda,
\end{equation}
where $\Omega_m$ is the matter energy density, and the dark energy density is $\Omega_\Lambda$.
Since $\Omega_m+\Omega_\Lambda=1$ for a flat $\Lambda$CDM model,
so we have only one independent parameter $\Omega_{m}$ in this model.
By fitting the $\Lambda$CDM model to the above data, we get the $1\sigma$ and $2\sigma$ constraints:
$a=-0.017_{-0.0718,-0.1415}^{+0.0717,+0.1416}$,
$b=-0.00013_{-0.0155,-0.0305}^{+0.0154,+0.0308}$, $\Omega_m=0.29_{-0.011,-0.022}^{+0.010,+0.022}$,
and $H_0=69.5_{-0.9,-1.7}^{+0.9,+1.8}$. The results show that Lorentz invariance is consistent with the data.
We show the results on $a$ and $b$ in Table \ref{all}.
The 1D probability distribution of each parameter and the 2D confidence contours for the parameters are shown in Fig.~\ref{L1}.

\begin{figure}
\begin{center}
\includegraphics[width=0.9\hsize]{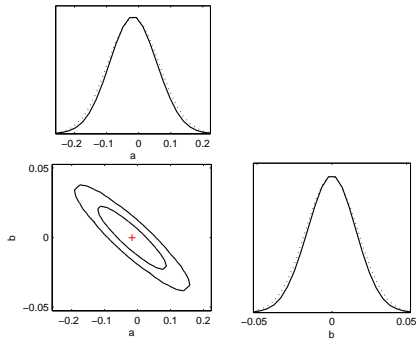}
\end{center}
\caption{The marginalized 1$\sigma$ and 2$\sigma$ contours and the distribution of the
LIV parameters for the $w$CDM model. The red cross represents the best-fit point. \label{w1}}
\end{figure}

For the flat $w$CDM model, the Friedmann equation is
\begin{equation}
h^2(z)= \Omega_m(1+z)^3 + \Omega_X(1+z)^{3(1+w)},
\end{equation}
where the dark energy density $\Omega_X=1-\Omega_m$.
Using the MCMC method, the marginalized $1\sigma$ and $2\sigma$ constraints on the model parameters are:
$a=-0.0168_{-0.0702,-0.1392}^{+0.0711,+0.1397}$,
$b=-0.00015_{-0.0154,-0.0304}^{+0.0153,+0.0303}$,
$\Omega_m=0.288_{-0.011,-0.021}^{+0.011,+0.022}$, $w=-1.05_{-0.045,-0.088}^{+0.04,+0.079}$, and
$H_0=70.2_{-1.1,-2.1}^{+1.1,+2.1}$. There is no evidence of LIV in $w$CDM model.
The results are shown in Table \ref{all}, and the marginalized plots are shown in Fig. \ref{w1}.

\begin{figure}
\begin{center}
\includegraphics[width=0.9\hsize]{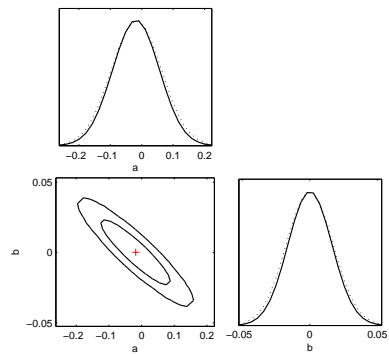}
\end{center}
\caption{ The marginalized 1$\sigma$ and 2$\sigma$ contours and the distribution of the LIV parameters for the CPL model.
The red cross represents the best-fit point. \label{CPL}}
\end{figure}

For the CPL model, the equation of state parameter is parameterized as $w(z)=w_0+w_1 z/(1+z)$ and
the Friedmann equation is
\begin{equation}
h^2(z)=\Omega_{m}(1+z)^3+\Omega_{X}(1+z)^{3(1+w_0+w_1)}\exp\left(-\frac{3w_1
z}{1+z}\right).
\end{equation}
In this model, $\Omega_{m}$, $w_0$, and $w_1$ are the model parameters.
Fitting the CPL model to the combined data, we
obtain the marginalized $1\sigma$ and $2\sigma$ constraints on the model parameters:
$a=-0.0183_{-0.0711,-0.1401}^{+0.0712,+0.14}$,
$b=0.00018_{-0.0155,-0.0301}^{+0.0159,+0.0306}$,
$\Omega_m=0.288_{-0.012,-0.022}^{+0.012,+0.024}$,
$w_0=-1.02_{-0.122,-0.237}^{+0.123,+0.246}$,
$w_1=-0.203_{-0.595,-1.281}^{+0.593,+1.053}$, and
$H_0=70.3_{-1.0,-2.57}^{+1.27,+2.45}$. We see no evidence of LIV in the CPL model.
The constraints on $a$ and $b$ are shown in Table \ref{all}, the marginalized probability distributions and the
$1\sigma$ and $2\sigma$ contours of $a$ and $b$ are shown in Fig. \ref{CPL}.

\begin{figure}
\begin{center}
\includegraphics[width=0.9\hsize]{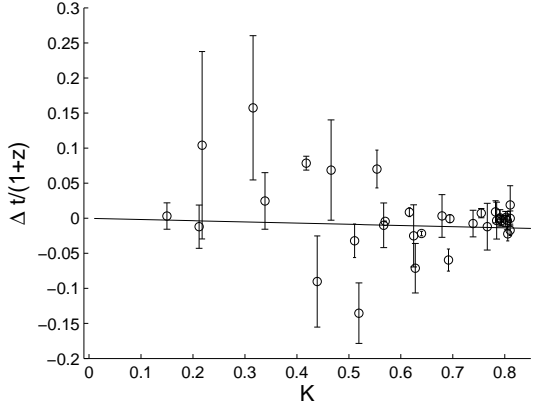}
\end{center}
\caption{The re-scaled spectral time-lags $\Delta t_{obs}/(1+z)$ versus $K(z)$ for the $\Lambda$CDM model with $\Omega_m=0.29$.
The solid line represents the best fit with the linear regression.
\label{figfour}}
\end{figure}

\begin{figure}
\begin{center}
\includegraphics[width=0.9\hsize]{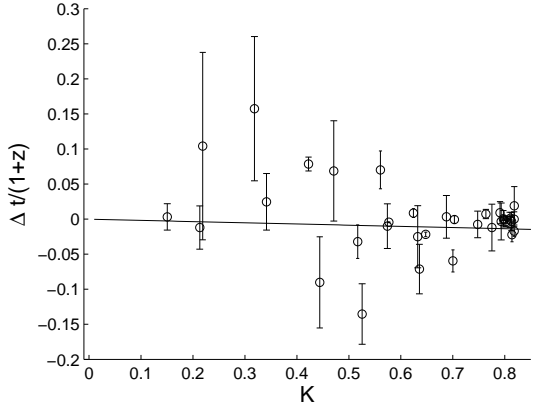}
\end{center}
\caption{The re-scaled spectral time-lags $\Delta t_{obs}/(1+z)$ versus $K(z)$ for the
$w$CDM model with $\Omega_m=0.288$ and $w=-1.05$.
The solid line represents the best fit with the linear regression.
\label{figfive}}
\end{figure}

\begin{figure}
\begin{center}
\includegraphics[width=0.9\hsize]{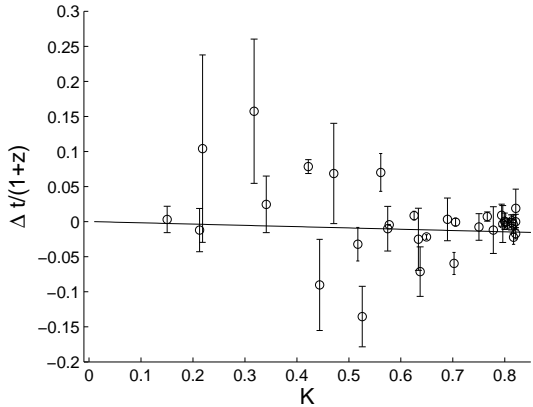}
\end{center}
\caption{The re-scaled spectral time-lags $\Delta t_{obs}/(1+z)$ versus $K(z)$ for the
CPL model with $\Omega_m=0.288$, $w_0=-1.02$ and $w_1=-0.203$. \label{figsix}}
\end{figure}

In the previous analysis, the model parameters take fixed values and a weak indication of LIV was found \citep{Ellis06,Biesiada09}.
Here we consider the possible degeneracies among all the parameters and take all the parameters as free parameters, we find
no evidence of LIV and all the results are consistent with each other. The conclusion is independent of the background model.
Although the uncertainties on the parameters $a$ and $b$ become larger due to more fitting parameters, the difference
on $a$ for different cosmological model is still relatively small. To better understand the effect of the background model,
we show the re-scaled spectral time lags $\Delta t_{obs}/(1+z)$ versus the $K(z)$ function for the three different
cosmological models in Figs. \ref{figfour}-\ref{figsix}, the data from GRBs is also shown in the figures.
Because the background cosmological model changes
the value of $K(z)$, so the value of $a$ also changes. Unlike the models considered in \citet{Biesiada09},
the $w$CDM model and CPL model considered here is close to the $\Lambda$CDM model as seen from the above best-fitting values
of $w$, $w_0$ and $w_1$, so our results are all consist with each other.

\begin{figure}
\begin{center}
\includegraphics[width=0.9\hsize]{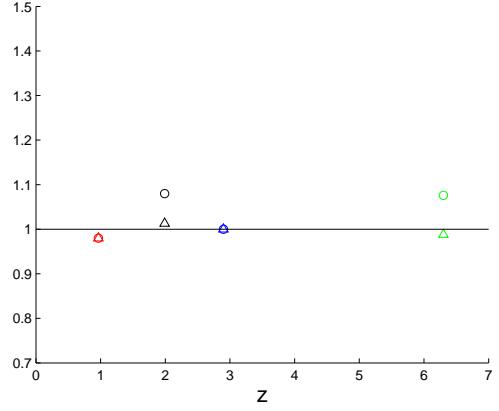}
\end{center}
\caption{The constraint on the slope $a$ for the $w$CDM and CPL models normalized to $\Lambda$CDM model
by using four sub-samples of GRBs. The GRBs are divided
into 4 groups with upper boundaries at $z=1.0$, $z=2.0$, $z=3.0$ and $z=6.3$.
The circles represent the CPL model and the triangles are for $w$CDM model. The red, black, blue and green
are for the fits to data with redshifts $z<1$, $z<2$, $z<3$ and $z>3$, respectively.
 \label{rsfGRBs}}
\end{figure}

The redshits of the GRBs span from $z=0.168$ to $z=6.29$, so we also consider the
effect of the redshift distribution on the results. We divide the GRBs
into 4 groups with upper boundaries at $z=1.0$, $z=2.0$, $z=3.0$ and $z=6.3$. The first group has 11 GRBs with redshifts $0<z<1.0$,
the second group has 21 GRBs with redshifts $0<z<2.0$, the third group has 27 GRBs with redshifts $z<3.0$ and the fourth group contains all 35 GRBs.
We then fit the cosmological models to each group of GRBs, and the constraints on
the slope $a$ normalized to the $\Lambda$CDM model are shown in Fig. \ref{rsfGRBs}.
The difference between different models are negligible in the first
and the third redshift groups. For both $w$CDM and the CPL models, the effects of redshift distributions
are similar although the deviation is a little larger for the CPL model.
The biggest contribution
comes from the GRBs  with redshifts in the range $1<z<2$ and $z>3$. As we discussed above,
the CPL model deviates more from $\Lambda$CDM than $w$CDM model does, so the value
of the slope $a$ for the $\Lambda$CDM model is more negative.
The reason of the redshift-dependence is due to distributions of the GRBs. In the first and third groups,
the average of the spectral time lags is close to zero. But we have more negative $\Delta t$ data in the redshift intervals
$1<z<2$ and $z>3$, therefore  we see bigger deviations for the data up to redshift 2 and for the whole data.

\section{Conclusions}\label{conclusions}

We test the LIV with the 35 GRBS, the Union2 SNeIa, the CMB and BAO data for three different cosmological models.
The slope $a$ and the intercept $b$
in the linear regression method as well as the model parameters are fitted to the combined data.
For the $\Lambda$CDM model, the marginalized $1\sigma$ and $2\sigma$ results are:
$a=-0.017_{-0.0718,-0.1415}^{+0.0717,+0.1416}$, and
$b=-0.00013_{-0.0155,-0.0305}^{+0.0154,+0.0308}$.
For the $w$CDM model, the marginalized $1\sigma$ and $2\sigma$ results are:
$a=-0.0168_{-0.0702,-0.1392}^{+0.0711,+0.1397}$, and
$b=-0.00015_{-0.0154,-0.0304}^{+0.0153,+0.0303}$.
For the CPL model, we obtain the marginalized $1\sigma$ and $2\sigma$ results
$a=-0.0183_{-0.0711,-0.1401}^{+0.0712,+0.14}$, and
$b=0.00018_{-0.0155,-0.0301}^{+0.0159,+0.0306}$. These results are also summarized in Table \ref{all}.
Because the slope $a$ is consistent with $a=0$ for all the three models, there is no evidence of LIV.
The results for all the three models are also consistent. Our conclusion is in conflict with previous result by \citet{Biesiada09}.
In their studies, the model parameters are fixed at certain values, so the degeneracies among parameters are neglected.
The dynamical dark energy models with the fixed parameter values
discussed in the previous work are very different from $\Lambda$CDM model, so they found
different results for different cosmological models.
In our analysis, we treat the model parameters as free parameters, and the model parameters are fitted with the latest SNeIa and BAO data.
The best fitted $w$CDM and CPL models are close to $\Lambda$CDM model.
Although the uncertainties on the model parameters become almost double compared with
the previous analysis \citep{Biesiada09} due to more fitting parameters, the differences between
different model are relatively small and all the results are consistent. Because the GRBs' data is not uniformly
distributed, especially in the redshift intervals $1<z<2$ and $z>3$,
we see the small differences between different models come from the GRBs at the redshift intervals $1<z<2$ and $z>3$.
In conclusion, our results show no evidence of LIV
in the three cosmological models.

The results are also in good agreement with the recent neutrino analysis,
which discussed LIV using two cascade neutrino events with energies around 1 PeV recently detected by IceCube \cite{Borriello13}.
Yet it is worth noting that experimental probes of LIV are limited by the scarcity of GRB data.
Other high-energy astrophysics experiments
such as the photon time-delay measurements from objects like Pulsars \citep{Kaaret99} and Active Galactic Nuclei(AGN) \citep{Albert08}
 may provide complementary probe of LIV effect.
Further studies are still needed to draw a more quantitative conclusion.

\section*{Acknowledgments}

This work was supported by the Ministry of Science and Technology National Basic Science Program (Project 973) under Grants Nos. 2012CB821804 and 2014CB845806, the National Natural Science Foundation of China under Grants Nos. 11073005, 11373014 and 11447213, the Fundamental Research Funds for the Central Universities and Scientific Research Foundation of Beijing Normal University, and China Postdoctoral Science Foundation under Grant No. 2014M550642. Y.G. is supported by the National Natural Science Foundation of China under grant Nos. 11175270 and 11475065, and the Program for New Century Excellent Talents in University under grant No. NCET-12-0205.  This work was also supported by Scientific and Technological Research Program of Chongqing Municipal Education Commission (Grant No. KJ130535), the Scientific Research Foundation For Doctor of Chongqing University of Posts and Telecommunications (A2013-25),HG acknowledges NASA NNX 13AH50G
.

\end{document}